\documentclass[12pt]{iopart}
\usepackage{epsf,iopams}
\begin{document}

\title{An embedding scheme for the Dirac equation}
\author{S. Crampin}
\address{Department of Physics, University of Bath, Bath BA2 7AY, United Kingdom}
\ead{s.crampin@bath.ac.uk}
\begin{abstract}
An embedding scheme is developed for the Dirac Hamiltonian $H$. 
Dividing space into 
regions I and II separated by surface $S$, an expression is derived for the
expectation value of $H$ which makes explicit reference to a 
trial function defined in I alone, with all details of region II replaced by 
an effective potential acting on $S$ and which is related to the Green 
function of region II. Stationary solutions provide approximations to 
the eigenstates of $H$ within I. The Green function for the
embedded Hamiltonian is equal to the Green function for the entire system
in region I.
Application of the method is illustrated for the problem of a hydrogen atom 
in a spherical cavity and an Au(001)/Ag/Au(001) sandwich structure
using basis sets that satisfy kinetic balance.
\end{abstract}

\submitto{\JPCM}


\section{Introduction}
\label{section:intro}
\begin{figure}
\bigskip

\epsfxsize= 80mm
\centerline{\epsffile{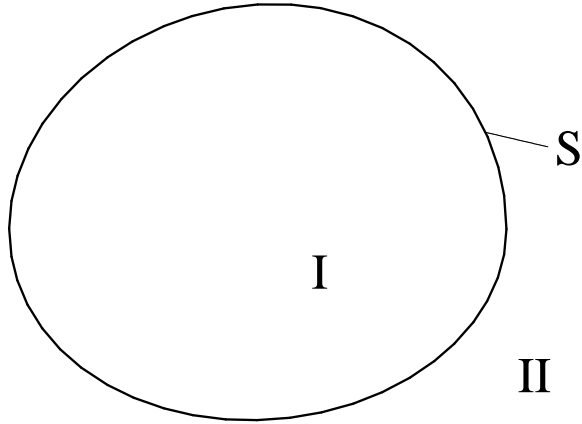}}
\bigskip

\caption{\label{fig:1}Schematic illustration of the embedded region I, the external region 
II, and the dividing surface $S$.}
\end{figure}

There are many problems concerning electronic structure where attention is 
focussed on a small region of a larger system, at surfaces or defects in
crystals being perhaps the most common. Let us call this region I, 
figure \ref{fig:1}, and the rest of the system region II. 
Although not of primary interest region II cannot be ignored, since in 
general the electron wave functions in I will be sensitive to the contents
of region II. Some time ago Inglesfield \cite{inglesfield81} derived 
an embedding scheme which enables the single-particle 
Schr\"odinger equation to be solved 
explicitly only in region I. The influence of region II is taken 
into account exactly by adding an energy-dependent non-local potential
to the Hamiltonian for region I, which constrains the solutions in I to match
onto solutions in II. This embedding method has been developed into a 
powerful tool most notably for surface electronic structure problems 
\cite{dev,crampin92,benesh94,ishida01} where it has found widespread application
especially to situations where an accurate description of the spectrum of 
electron states is necessary. Examples include studies of 
image states \cite{app1}, surface states at metals
surfaces \cite{app2}, static and dynamic screening \cite{app5},
atomic adsorption and scattering at surfaces \cite{trioni96}, 
studies of surface optical response \cite{optical} and field emission \cite{ohwaki03}. Recent applications to transport problems have also been
described \cite{transport}. For a review of the embedding method see 
Inglesfield \cite{inglesfield01}.

In the case of materials containing heavier elements, relativistic effects
can be significant \cite{nato} and lead to important deviations from 
the electronic 
structure as predicted by the Schr\"odinger equation -- shifts in
inner core levels of 5d elements are typically several 100 or 1000 eV, 
valence bands shifts are on the eV scale and spin orbit splitting is 
often measured in tenths of eV. Even ignoring the concomitant
changes in electron wave functions these shifts can reorder levels and
so affect calculated densities, fundamental to the determination of 
ground state properties within the density functional framework 
\cite{macdonald}.
For this reason most
of the conventional electronic structure techniques developed for accurately
solving the single-particle Schr\"odinger equation in solids have subsequently
been modified to deal with the Dirac equation, including the relativistic 
augmented plane wave method \cite{rapw,theileis00}, 
relativistic linear muffin-tin orbital 
method \cite{rlmto}, 
relativistic augmented spherical wave method 
\cite{rasw} and the
relativistic multiple-scattering method \cite{rmst}, and each has subsequently 
been used in studying a diverse range of problems. The last method 
alone has formed the basis of calculations of photoemission \cite{pe}, 
magnetocrystalline anisotropy \cite{anis}, hyperfine interactions \cite{hf} 
and magnetotransport \cite{mt} amongst other topics. 

Inglesfield's embedding method has particular advantages that encourage
its extension to the relativistic case. It permits the inclusion of extended
substrates for surface and interface calculations, enables the study of
isolated point defects in solids and being a basis set technique is
highly flexible and permits full-potential studies with relative ease.
At surfaces extended substrates (as against the use of the supercell
or thin-film approximation in which the crystal is approximated by a 
small number of layers, typically 5-7) enable the proper distinction 
between surface states, resonances and the continuum of bulk states 
\cite{app2}. The behaviour  of the W(110) surface \cite{rotenberg99}
where the addition of half a monolayer of Li is observed to 
{\it increase} the spin-orbit splitting
of a surface state by $\sim 0.5$ eV (resulting in Fermi surface crossings 
separated by $\sim 20$\% of the Brillouin zone dimension) typifies
a type of problem a relativistic embedding scheme could address.
Indeed each of the topics mentioned at the end of the previous paragraph
are relevant at surfaces and/or interfaces, and could be usefully investigated 
within a relativistic embedding framework.

In this paper we develop an embedding scheme for the Dirac 
equation that 
parallels Inglesfield's scheme for the Schr\"odinger equation. 
Inglesfield's starting point is the expectation value 
of the Hamiltonian using 
a trial wave function which is continuous in amplitude but discontinuous in 
derivative across the surface $S$ separating I and II. The first order nature
of the Dirac equation precludes the use of a similar trial function. Instead,
in the following section we use a trial function in which the large component 
is continuous and the small component discontinuous across $S$. Continuity
in the small component is restored when the resulting equations are solved
exactly. Using the
Green function for region II we are able to derive an expression for the
expectation value purely in terms of the trial function in I. In section
\ref{section:app} the application of the method is illustrated by calculating
the eigenstates of a hydrogen atom within a cavity and in section
\ref{section:green} we determine the Green function for the embedded region.
Section \ref{section:monolayer} briefly illustrates the method applied to a
sandwich structure where relativistic effects are marked.
We conclude with a brief summary and discussion.

\section{Embedding scheme}
\label{section:emb}
\begin{figure}
\bigskip

\epsfxsize= 60mm
\centerline{\epsffile{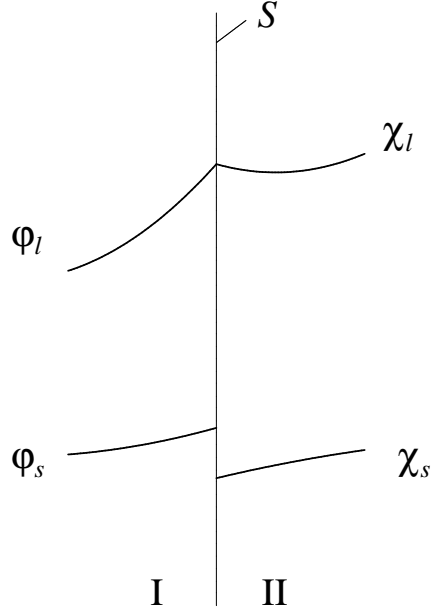}}
\bigskip

\caption{\label{fig:2}The large component of the trial function is defined to be 
continuous across the surface $S$ dividing regions I and II, but the
small component can be discontinuous.}
\end{figure}

In this section we consider region I joined onto region II 
(figure \ref{fig:1}), and derive a 
variational principle for a trial wave function $\varphi$ defined
explicitly only within region I. We are primarily interested in the positive
energy solutions of the Dirac equation \cite{rqm}, and so 
we refer to the upper and lower
spinors of the Dirac bi-spinor solutions as the large and small components
of the wave function respectively.
We notionally extend $\varphi$ into II as $\chi$, an exact solution
of the Dirac equation at some energy $w$, with the large
components of $\varphi$ and $\chi$ ($\varphi_{\rm l}$ and $\chi_{\rm l}$)
matching on the surface $S$ separating I
and II, but with no constraint upon the small components 
($\varphi_{\rm s}$ and $\chi_{\rm s}$), figure \ref{fig:2}. 
The expectation value for the energy $W$ is then
\begin{equation}
W=\frac{\langle\varphi|\widehat{H}|\varphi\rangle_{\mathrm{I}}+
\langle\chi|\widehat{H}|\chi\rangle_{\mathrm{II}}
+\rmi c\hbar \int_S \rmd\bi{r}_S \cdot \varphi_{\rm l}^\dag \bsigma
\left[\varphi_{\rm s}-\chi_{\rm s}\right]
}{\langle\varphi|\varphi\rangle_{\mathrm{I}}+
\langle\chi|\chi\rangle_{\mathrm{II}}}
\label{eqn:exp}
\end{equation}
where $\widehat{H}=c\balpha\cdot\widehat{\bi{p}}+\beta mc^2+V$.
(For clarity we omit the interaction 
$\propto \beta\bsigma\cdot\bi{B}$ which appears in the
relativistic density functional theory \cite{macdonald} neglecting orbital 
and displacement currents, where $\bi{B}$ 
is a ``spin-only'' effective  
magnetic field containing an external and exchange-correlation contribution.
Its inclusion has no consequences for the derivation.)
The first two terms in the numerator are the expectation value of the
Hamiltonian through regions I and II, and the third the
contribution due to the discontinuity in the small component of the
wave function on $S$ (in this and the following, surface normals are 
directed from I to II).

We eliminate reference to $\chi$ by introducing two relations. Firstly, 
for $\bi{r}\in\mathrm{II}$, $\chi$ satisfies the Dirac equation at energy 
$w$
\begin{equation}
-\rmi c\hbar \balpha\cdot\bnabla\chi+
\left[\beta mc^2 +V-w\right]\chi=0
\label{eqn:dirII}
\end{equation}
and differentiating with respect to $w$ the energy derivative of $\chi$,
$\dot{\chi}=\partial\chi/\partial w$,
satisfies
\begin{equation}
-\rmi c\hbar \balpha\cdot\bnabla\dot{\chi}+
\left[\beta mc^2 +V-w\right]\dot{\chi}=\chi\qquad \bi{r}\in\mathrm{II}.
\label{eqn:eder}
\end{equation}
Multiplying the Hermitian conjugate of the first equation by $\dot\chi$ 
from the right, multiplying the second from the left by $\chi^\dag$,
subtracting and integrating over region II gives a relation between 
the normalisation of $\chi$ in II and the amplitude on $S$:
\begin{equation}
\langle\chi|\chi\rangle_{\mathrm{II}}=\rmi c\hbar \int_S \rmd\bi{r}_S
\cdot \chi_{\rm l}^\dag\bsigma\,\dot{\chi}_{\rm s}.
\label{eqn:norm}
\end{equation}
We have assumed that $\chi$ vanishes sufficiently strongly at infinity.

For the second relation we introduce the Green function (resolvant)
$G(\bi{r},\bi{r}';w)$ corresponding to equation (\ref{eqn:dirII}):
\begin{equation}
-\rmi c\hbar \balpha\cdot\bnabla G+
\left[\beta mc^2 +V-w\right]G=-\delta(\bi{r}-\bi{r}') 
\qquad \bi{r},\bi{r}'\in \mathrm{II}.
\label{eqn:gf}
\end{equation}
Multiplying the Hermitian conjugate of this equation by $\chi$ from the right,
and subtracting $G^\dag$ times equation (\ref{eqn:dirII}), integrating over
region II and then using the reciprocity of the Green function 
gives
\begin{equation}
\chi(\bi{r})=\rmi c\hbar \int_S \rmd\bi{r}'_S\cdot G(\bi{r},\bi{r}_S';w)
\balpha\,\chi(\bi{r}_S') \qquad \bi{r}\in \mathrm{II}.
\label{eqn:chis}
\end{equation}
We see that the Green function relates the amplitude of the wave function
on $S$ to the amplitude at any point within II. In particular, we can obtain
a relation between the large and small components of $\chi$ on S. Writing
the $4\times 4$ Green function as
\begin{equation}
G(\bi{r},\bi{r}';w)= 
\left( 
\begin{array}{cc}
G_{\rm ll}(\bi{r},\bi{r}';w) & G_{\rm ls}(\bi{r},\bi{r}';w) \\
G_{\rm sl}(\bi{r},\bi{r}';w) & G_{\rm ss}(\bi{r},\bi{r}';w) 
\end{array} 
\right)
\label{eqn:gre}
\end{equation}
where each entry is a $2 \times 2$ matrix, 
substituting into equation (\ref{eqn:chis}),
and rearranging the two equations coupling the small and large components
of $\chi$ gives
\begin{equation}
\chi_{\rm s}(\bi{r}_S)={\rmi }c\hbar \int_S \rmd\bi{r}'_S \cdot 
\Gamma(\bi{r}_S,\bi{r}'_S;w)\bsigma\,\chi_{\rm l}(\bi{r}'_S)
\label{eqn:cgc}
\end{equation}
where
\begin{equation}
\fl
\Gamma(\bi{r}_S,\bi{r}'_S;w)=G_{\rm ss}(\bi{r}_S,\bi{r}'_S;w) 
+{\rmi }c\hbar \int_S \rmd\bi{r}''_S \cdot G_{\rm sl}(\bi{r}_S,\bi{r
}''_S;w)\bsigma\,\Gamma(\bi{r}''_S,\bi{r}'_S;w)
\label{eqn:gam}
\end{equation}
It follows from  (\ref{eqn:chis}) that the Green functions in (\ref{eqn:gam})
are the limiting forms of $G(\bi{r},\bi{r}_S;w)$ as 
$\bi{r}\rightarrow \bi{r}_S$
from within II.

Equations (\ref{eqn:norm}) and (\ref{eqn:cgc}) are the desired results that 
enable us to express the expectation value $W$ in (\ref{eqn:exp}) in terms
of $\varphi$ alone. After substitution and use of the
continuity of the large components $\chi_{\rm l}=\varphi_{\rm l}$ on $S$ we obtain
\begin{equation}
W=\frac{\langle\varphi|\widehat{H}|\varphi\rangle_{\mathrm{I}}+
{\rmi }c\hbar \int_S \rmd\bi{r}_S \cdot \varphi_{\rm l}^\dag \bsigma
\left[\varphi_{\rm s}-{\rmi }c\hbar \int_S \rmd\bi{r}'_S\cdot \left\{\Gamma-w
\dot{\Gamma}\right\}\bsigma\varphi_{\rm l}\right]
}{
\langle\varphi|\varphi\rangle_{\mathrm{I}}-
c^2\hbar^2 \int_S \rmd\bi{r}_S\cdot \varphi_{\rm l}^\dag \bsigma
\int_S \rmd\bi{r}'_S \cdot \dot{\Gamma} \bsigma\, \varphi_{\rm l}
}.
\label{eqn:exp2}
\end{equation}
This is an expression for the expectation value of the energy $W$,
given purely in terms of the trial function $\varphi$ in region I and on the 
surface $S$, with all details of region II entering via $\Gamma$ and its energy
derivative. Following the convention in the non-relativistic embedding 
scheme we shall refer to $\Gamma$ as the embedding potential.

To see what this variational principle means in practice, we consider 
variations in $\varphi^\dag$, whereby
\begin{eqnarray}
\fl
\delta W=
\frac{\langle\delta\varphi|\widehat{H}-W|\varphi\rangle_{\mathrm{I}}+
{\rmi }c\hbar\int_S \rmd\bi{r}_S \cdot \delta\varphi_{\rm l}^\dag \bsigma
\left[\varphi_{\rm s}-{\rmi }c\hbar\int_S \rmd\bi{r}'_S\cdot \left\{\Gamma+(W-w)
\dot{\Gamma}\right\}\bsigma\varphi_{\rm l}\right]
}{
\langle\varphi|\varphi\rangle_{\mathrm{I}}-
c^2\hbar^2\int_S \rmd\bi{r}_S\cdot \varphi_{\rm l}^\dag \bsigma
\int_S \rmd\bi{r}'_S \cdot \dot{\Gamma} \bsigma\, \varphi_{\rm l}
},\nonumber\\
\label{eqn:de1}
\end{eqnarray}
so that solutions $\varphi$ stationary with respect to arbitrary variations 
$\delta\varphi$ satisfy
\numparts
\begin{eqnarray}
\fl
\widehat{H}\varphi&=&W\varphi\qquad \bi{r}\in{\mathrm{I}} \\ 
\fl
\varphi_{\rm s}(\bi{r}_S)&=&{\rmi }c\hbar\int_S \rmd\bi{r}'_S\cdot\left\{
\Gamma(\bi{r}_S, \bi{r}'_S;w)+
(W-w)\dot{\Gamma}(\bi{r}_S, \bi{r}'_S;w)\right\}\bsigma
\varphi_{\rm l}(\bi{r}'_S).
\label{eqn:de2}
\end{eqnarray}
\endnumparts
The first expression indicates $\varphi$ is a solution of the Dirac 
equation at energy $W$ in region I. Comparing the second with (\ref{eqn:cgc}) shows
that $\varphi$ also possesses the
correct relationship between large and small components on $S$, 
the surface separating I and II, to match onto solutions in II.
The term $(W-w)\dot{\Gamma}(w)$ provides a first order correction to 
$\Gamma(w)$ so that the boundary condition is appropriate for
energy $W$.

In practice expression (\ref{eqn:exp2}) may be used to obtain solutions of 
the Dirac Hamiltonian by inserting a suitably parameterised trial function
and varying the parameters to obtain a stationary solution. This is
conveniently achieved by 
expanding the trial solution in a finite basis of separate large and small 
component spinors
\begin{equation}
\fl
\varphi(\bi{r})=
\sum_{n=1}^{N_{\rm l}} a_{{\rm l},n} 
\left[\begin{array}{c}\psi_{{\rm l},n}(\bi{r})\\0\end{array}\right]+
\sum_{n=1}^{N_{\rm s}} a_{{\rm s},n} 
\left[\begin{array}{c}0\\ \psi_{{\rm s},n}(\bi{r})\end{array}\right]
=\left[\begin{array}{cc}
\bpsi_{\rm l}(\bi{r}) & 0 \\ 0 & \bpsi_{\rm s}(\bi{r})
\end{array}\right]
\left[
\begin{array}{c}
\bi{a}_{\rm l} \\ \bi{a}_{\rm s}
\end{array}
\right].
\end{equation}
The matrix in the final expression is $4$ by $N_{\rm l}+N_{\rm s}$, 
and the column
vector contains the $N_{\rm l}+N_{\rm s}$ coefficients. Substituting into 
(\ref{eqn:exp2}) we find
states $\phi$ that are stationary with respect to variations in 
the expansion
coefficients $\{a_{{\rm l},n}, a_{{\rm s},n}\}$ are then given by the eigenstates of a
generalised eigenvalue problem of the form
\begin{equation}
\left[
\begin{array}{cc}
{H}_{\rm ll} & {H}_{\rm ls} \\ {H}_{\rm sl} & {H}_{\rm ss}
\end{array}
\right]
\left[
\begin{array}{c}
\bi{a}_{\rm l} \\ \bi{a}_{\rm s}
\end{array}
\right]
=W
\left[
\begin{array}{cc}
{O}_{\rm ll} & 0 \\ 0 & {O}_{\rm ss}
\end{array}
\right]
\left[
\begin{array}{c}
\bi{a}_{\rm l} \\ \bi{a}_{\rm s}
\end{array}
\right],
\label{eqn:a1}
\end{equation}
where
\numparts
\begin{eqnarray}
\fl
\left[{H}_{\rm ll}\right]_{nn'}=
\int_I \psi_{{\rm l},n}^\dag(\bi{r}) \left(V(\bi{r})+mc^2 \right) 
\psi_{{\rm l},n'}(\bi{r}) \rmd\bi{r} \nonumber \\
\fl
\phantom{\left[{H}_{\rm ll}\right]_{nn'}=}{}+c^2\hbar^2\!\int_S \rmd\bi{r}_S\cdot 
\psi_{{\rm l},n}^\dag(\bi{r}_S)\bsigma\!
\int_S \rmd\bi{r}'_S \!\cdot\! 
\left[ \Gamma(\bi{r}_S,\bi{r}'_S;w)\! - \! w
\dot{\Gamma}(\bi{r}_S,\bi{r}'_S;w)\right]
\bsigma\psi_{{\rm l},n'}(\bi{r}'_S) \label{eqn:h2}\\
\fl
\left[{H}_{\rm ls}\right]_{nn'}=
\int_I \psi_{{\rm l},n}^\dag(\bi{r}) c\bsigma\cdot \widehat{\bi{p}} 
\psi_{{\rm s},n'}(\bi{r}) \rmd\bi{r}+\rmi c\hbar\int_S \rmd\bi{r}_S\cdot 
\psi_{{\rm l},n}^\dag(\bi{r}_S)\bsigma\psi_{{\rm s},n'}(\bi{r}_S)\label{eqn:h3}\\
\fl
\left[{H}_{\rm sl}\right]_{nn'}=
\int_I \psi_{{\rm s},n}^\dag(\bi{r}) c\bsigma\cdot\widehat{\bi{p} }
\psi_{{\rm l},n'}(\bi{r}) \rmd\bi{r} \label{eqn:h4}\\
\fl
\left[{H}_{\rm ss}\right]_{nn'}=
\int_I \psi_{{\rm s},n}^\dag(\bi{r}) \left(V(\bi{r})-mc^2 \right) 
\psi_{{\rm s},n'}(\bi{r}) \rmd\bi{r} \label{eqn:h5}\\
\fl
\left[{O}_{\rm ll}\right]_{nn'}=
\int_I \psi_{{\rm l},n}^\dag(\bi{r}) 
\psi_{{\rm s},n'}(\bi{r}) \rmd\bi{r}\nonumber\\
\fl
\phantom{\left[{O}_{\rm ll}\right]_{nn'}=}{}
-c^2\hbar^2\int_S \rmd\bi{r}_S\cdot 
\psi_{{\rm l},n}^\dag(\bi{r}_S)\bsigma
\int_S \rmd\bi{r}'_S \cdot 
\dot{\Gamma}(\bi{r}_S,\bi{r}'_S;w)
\bsigma\psi_{{\rm l},n'}(\bi{r}'_S) \label{eqn:h6}\\
\fl
\left[{O}_{\rm ss}\right]_{nn'}=
\int_I \psi_{{\rm s},n}^\dag(\bi{r}) \psi_{{\rm s},n'}(\bi{r}) 
\rmd\bi{r}.\label{eqn:h7}
\end{eqnarray}
\endnumparts
Of course the spectrum of the Dirac Hamiltonian is unbounded below, and 
care must be taken to prevent solutions collapsing to negative energies. 
This can be avoided through the use of a kinetically balanced basis 
\cite{stanton} in which there is a one-to-one
relationship between large and small component spinors, $N_{\rm s}=N_{\rm l}=N$, and
where the small component spinors are given by
\begin{equation}
\psi_{{\rm s},n}(\bi{r})=\bsigma\cdot\widehat{\bi{p}}\ \psi_{{\rm l},n}(\bi{r}).
\label{eqn:kinbal}
\end{equation}
The upper half of the spectrum of the $2N$ eigenstates of (\ref{eqn:a1})
then provide approximations to the spectrum of electronic states.

\section{Model application}
\label{section:app}

\begin{figure}
\bigskip

\epsfxsize= 80mm
\centerline{\epsffile{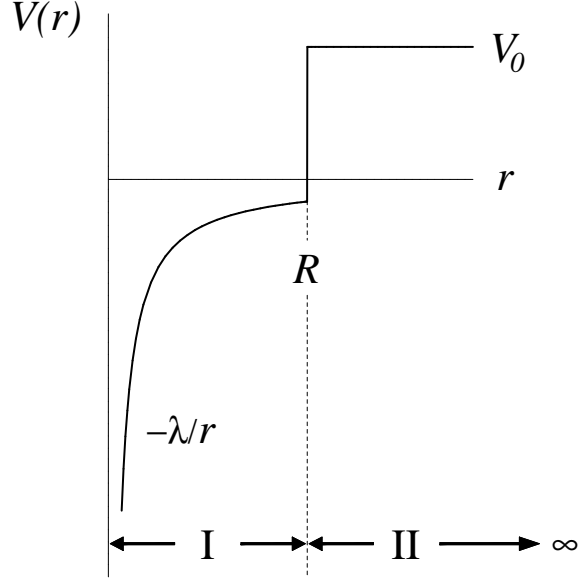}}
\bigskip

\caption{\label{fig:3}The model potential used to illustrate the relativistic embedding 
scheme.}
\end{figure}

To illustrate the application of the relativistic embedding scheme
we consider a model problem of a hydrogen atom within a spherical cavity,
finding bound states of the Dirac equation corresponding to the potential 
illustrated in figure \ref{fig:3}:
\[
V(\bi{r})=\left\{ 
\begin{array}{cll}
-{\lambda/ r} && r \le R \\
V_0 && r > R
\end{array}
\right.
\]
where $\lambda=e^2/(4\pi\epsilon_0)$ and $V_0>0$. We choose this model as the bound states 
may also be found straightforwardly by alternative methods.
Region I, the region to be treated 
explicitly, is the sphere of radius $R$ centered on $r=0$. The external
region II where $V(\bi{r})=V_0$ is replaced by an embedding potential
acting on the surface of the sphere. The value of the embedding potential
is most readily evaluated from equation (\ref{eqn:cgc}). 
A general solution to the Dirac equation at some energy $w$ in region II and 
satisfying the appropriate boundary conditions is \cite{rqm}
\begin{equation}
\chi(\bi{r})=
\left(
\begin{array}{c}
\chi_{\rm l}(\bi{r}) \\ \chi_{\rm s}(\bi{r})
\end{array}
\right)
=
\sum_\Lambda a_\Lambda \sqrt{\frac{\pi}{2 k r }}
\left(
\begin{array}{c}
K_{\ell+\frac{1}{2}}(kr)\Omega_{\Lambda}(\bi{r}) \\
-i\gamma_\kappa K_{\bar{\ell}+\frac{1}{2}}(kr)\Omega_{\bar{\Lambda}}(\bi{r}) \\
\end{array}
\right)
\label{eqn:a2}
\end{equation}
where $\Lambda=(\kappa,\mu)$, $\bar{\Lambda}=(-\kappa,\mu)$,
$\Omega_\Lambda(\bi{r})$ a spin-angular function,
$K_{n+\frac{1}{2}}(z)$ a modified spherical Bessel 
function of the third kind \cite{abram}, 
$c\hbar k=\sqrt{m^2 c^4 - (w-V_0)^2}$, $\gamma_\kappa=\hbar c k/(w-V_0+mc^2)$
and
\begin{equation}
\ell=\left\{
\begin{array}{cll}
\kappa && \kappa > 0 \\ -(\kappa+1) && \kappa < 0
\end{array}, \right. \qquad  \bar{\ell}=\ell-\frac{\kappa}{|\kappa|}.
\end{equation}
The spherical symmetry of region II means the the 
embedding potential $\Gamma$ may be expanded on $S$ as
\begin{equation}
\Gamma(\bi{r}_S,\bi{r}'_S;w)=\sum_{\Lambda}
\Gamma_{\kappa}(w)\Omega_{\bar{\Lambda}}(\bi{r}_S)
\Omega_{\bar{\Lambda}}^\dag(\bi{r}'_S)
\label{eqn:a2b}
\end{equation}
and substituting (\ref{eqn:a2b}) and (\ref{eqn:a2}) into (\ref{eqn:cgc}) leads 
to
\begin{equation}
\Gamma_\kappa(w)=\frac{\gamma_\kappa}{c\hbar R^2}
\frac{K_{\bar{\ell}+\frac{1}{2}}(kR)}{K_{\ell+\frac{1}{2}}(kR)}.
\label{eqn:a4}
\end{equation}
Using (\ref{eqn:gam}) with the Green function for constant potential ($w<V_0, r>r'$)
\begin{eqnarray}
\fl
G(\bi{r},\bi{r}';w)= -\frac{(w-V_0+mc^2)}{c^2\hbar^2 r}\nonumber\\
\times\sum_\Lambda
\left(
\begin{array}{c}
K_{\ell+\frac{1}{2}}(kr)\Omega_{\Lambda}(\bi{r}) \\
-i\gamma_\kappa K_{\bar{\ell}+\frac{1}{2}}(kr)\Omega_{\bar{\Lambda}}(\bi{r}) \\
\end{array}
\right)
\left(
\begin{array}{c}
I_{\ell+\frac{1}{2}}(kr')\Omega_{\Lambda}(\bi{r}') \\
i\gamma_\kappa I_{\bar{\ell}+\frac{1}{2}}(kr')\Omega_{\bar{\Lambda}}(\bi{r}')
\end{array}
\right)^\dag
\end{eqnarray}
gives the same result but after rather more involved manipulations. 
$I_{n+\frac{1}{2}}(z)$ is a modified spherical Bessel function of 
the first kind.

Because of the spherical symmetry we can determine separately states 
with a given angular character $\Lambda$.
Using as a basis set for the large component spinors
\begin{equation}
\psi^{(\Lambda)}_{{\rm l},n}(\bi{r})=\frac{1}{r} g_n(r)\Omega_{\Lambda}(\bi{r}),
\qquad g_n(r)=r^n e^{-r}
\label{eqn:bas1}
\end{equation}
so that the small component spinors ensuring kinetic balance are
\begin{equation}
\psi^{(\Lambda)}_{{\rm s},n}(\bi{r})=
\frac{i}{r}f_{n\kappa}(r)\Omega_{\bar{\Lambda}}(\bi{r}),
\qquad f_{n\kappa}(r)=\hbar \left[ (n+\kappa)r^{n-1}-r^n\right] e^{-r},
\label{eqn:bas2}
\end{equation}
the matrix elements become
\numparts
\begin{eqnarray}
\fl
\left[{H}^{(\Lambda)}_{\rm ll}\right]_{nn'}=
\int_0^R g_{n}(r)\left[-\frac{\lambda}{r}+mc^2\right] g_{n'}(r) \rmd r
\nonumber\\
+\hbar^2c^2R^2g_{n}(R)g_{n'}(R)
\left[\Gamma_\kappa(w)-w\dot{\Gamma}_\kappa(w)\right] \\
\fl
\left[{H}^{(\Lambda)}_{\rm ss}\right]_{nn'}=
\int_0^Rf_{n\kappa}(r)\left[-\frac{\lambda}{r}-mc^2\right]f_{n'\kappa}(r) 
\rmd r\\
\fl
\left[{H}^{(\Lambda)}_{\rm ls}\right]_{nn'}=-\hbar c
\int_0^R g_{n}(r)\left[\frac{\rmd f_{n'\kappa}(r)}{\rmd r}-
\frac{\kappa}{r}f_{n'\kappa}(r)\right] \rmd r +
\hbar c g_{n}(R)f_{n'\kappa}(R)\\
\fl
\left[{H}^{(\Lambda)}_{\rm sl}\right]_{nn'}=\hbar c
\int_0^R f_{n\kappa}(r)\left[\frac{\rmd g_{n'}(r)}{\rmd r}+
\frac{\kappa}{r}g_{n'}(r)\right] \rmd r \\
\fl
\left[{O}^{(\Lambda)}_{\rm ll}\right]_{nn'}=
\int_0^R g_n(r)g_{n'}(r) \rmd r
-\hbar^2c^2R^2g_n(R)g_{n'}(R)\dot{\Gamma}_\kappa(w) \\
\fl
\left[{O}^{(\Lambda)}_{\rm ss}\right]_{nn'}=
\int_0^R f_{n\kappa}(r)f_{n'\kappa}(r) \rmd r
\end{eqnarray}
\endnumparts
The eigenvalues only depend upon the quantum number $\kappa$.
In table \ref{table:1} the lowest two eigenvalues of $\kappa=-1$ symmetry
(corresponding to the $1s_{1/2}$ and $2s_{1/2}$ of free hydrogen) are 
shown as a function of basis set size and for different values of the energy
$w$ at which the embedding potential is evaluated, for the case $R=3$, 
$V_0=10$. For comparison also
given are the values found by
matching the external solution (\ref{eqn:a2}) to the regular internal
solution, which can be expressed in terms of confluent hypergeometric
functions\cite{rqm}. 
For a given fixed $w$ the eigenvalues converge from above to 
values that are equal or above the exact values. The further $w$ lies 
from the eigenvalue, the larger the difference between the limiting value for
large basis sets and the correct value. However, the influence of
the $\dot{\Gamma}$ terms in (\ref{eqn:exp2}) means the error is relatively 
small.
When $w=mc^2$, the lowest eigenvalue found with $N_{\rm l}=8$ is $-0.445\,5488$ Ha
and in error by only 0.000\,0044 Ha, a factor $10^5$ smaller than the error 
in $w$.

\begin{table}
\caption{\label{table:1}Lowest two electron-like eigenvalues $E=W-mc^2$ 
of $s_{1/2}$ symmetry of a 
hydrogen atom confined to a spherical cavity with radius $R=3$ and 
confining potential $V=10$, obtained using embedding potentials at different
trial energies $w$. $w=W$ indicates the solution has been iterated to
ensure $w$ coincides with the eigenvalue. The exact eigenvalues are those
found by matching internal and external solutions at $R$. We use atomic units
$e^2=\hbar=m=1$ and $c=137.035\,999\,76$ so that the corresponding free atom 
eigenvalues are $-0.500\,0067, -0.125\,0021$}
\begin{indented}
\item[]\begin{tabular}{@{}cccc}
\br
$N_{\rm l}$ & $w=mc^2-0.5$ & $w=mc^2$ &  $ w=W$ \\
\mr
2 & -0.411\,1620,\ 1.698\,0995 & -0.411\,1527, 1.694\,9300 & -0.411\,1624, 1.689\,6482 \\
4 & -0.445\,1482,\ 0.912\,9418 & -0.445\,1439, 0.912\,6817 & -0.439\,6204, 0.971\,4775 \\
6 & -0.445\,5519,\ 0.891\,4789 & -0.445\,5477, 0.891\,2219 & -0.445\,5520, 0.891\,0268 \\
8 & -0.445\,5532,\ 0.891\,2708 & -0.445\,5488, 0.891\,0141 & -0.445\,5532, 0.890\,8194 \\
``exact'' & -0.445\,5532, 0.890\,8194 & -0.445\,5532, 0.890\,8194 
& -0.445\,5532, 0.890\,8194\\
\br
\end{tabular}
\end{indented}
\end{table}

Differentiating (\ref{eqn:exp2}) with respect to the trial energy $w$
shows the expectation value is stationary at $w=W$. In this case $W$ is
given by the solutions of 
\begin{equation}
W=\frac{\langle\varphi|\widehat{H}|\varphi\rangle_{\mathrm{I}}+
{\rmi }c\hbar \int_S \rmd\bi{r}_S \cdot \varphi_{\rm l}^\dag \bsigma
\left[\varphi_{\rm s}-{\rmi }c\hbar \int_S \rmd\bi{r}'_S\cdot \Gamma(W)
\bsigma\varphi_{\rm l}\right]
}{\langle\varphi|\varphi\rangle_{\mathrm{I}}
}.
\label{eqn:exp3}
\end{equation}
Eigenfunctions $\varphi$ solving this equation satisfy the Dirac equation
within I and the relationship between small and large components on $S$
(\ref{eqn:de2}) is exact. 
The final column in table \ref{table:1} shows the lowest two positive 
energy eigenvalues of (\ref{eqn:exp3}), again as a function of basis set size. 
The eigenvalues again converge from above, and by $N_{\rm l}=8$
reproduce the exact values by at least 7 significant figures.
It is worth noting that with this particular basis set increasing 
$N_{\rm l}$ much 
further leads to some numerical difficulties due to overcompleteness. 
For more accurate work a more suitable basis set should be used. It should
also be noted that conventional finite basis set
calculations using a basis satisfying kinetic balance can given eigenvalues
that lie below exact limiting values by an amount of order 
$1/c^4$ \cite{stanton}, and similar behaviour is expected in this 
embedding scheme.

\section{Green function}
\label{section:green}

Most practical applications of the Schr\"odinger embedding scheme have 
actually used the Green function of the embedded system. This is a more 
convenient quantity when dealing with systems where the spectrum is
continuous, such as at surfaces or defects in solids. 
We therefore consider the Green function for
the embedded Dirac system.

Differentiating (\ref{eqn:exp2}) with respect to $w$ shows $W$ is stationary 
when $w=W$, as would be expected. In this case stationary solutions
satisfy the embedded Dirac equation
\begin{equation}
\widehat{H}\varphi(\bi{r})+
\int_I \rmd\bi{r}' \Delta(\bi{r},\bi{r}';W) \varphi(\bi{r}')
-W\varphi(\bi{r})=0 \qquad \bi{r}\in I
\label{eqn:gf1}
\end{equation}
where, introducing $\sigma_S(\bi{r}_S)$, the component of 
$\bsigma$ in the direction normal to the surface
$S$ (from I to II) at $\bi{r}_S$,
the additional term $\Delta$ enforcing the embedding is
\begin{eqnarray}
\fl
\Delta(\bi{r},\bi{r}';w)=
\int_S \rmd\bi{r}_S 
\delta(\bi{r}-\bi{r}_S)
\int_S \rmd\bi{r}'_S 
\delta(\bi{r}'-\bi{r}'_S)
\nonumber \\
\times
\left[
\begin{array}{cc}
c^2\hbar^2\sigma_S(\bi{r}_S)\Gamma(\bi{r}_S,\bi{r}'_S;w)\sigma_S(\bi{r}'_S)
& \rmi c\hbar \sigma_S(\bi{r}_S)\delta(\bi{r}_S-\bi{r}'_S) \\
0 & 0
\end{array}
\right].
\label{eqn:gf2}
\end{eqnarray}
The corresponding Green function satisfies 
\begin{equation}
\fl
\widehat{H}G(\bi{r},\bi{r}';W)
+\int_I \rmd\bi{r}'' \Delta(\bi{r},\bi{r}'';W) 
G(\bi{r}'',\bi{r}';W) 
-WG(\bi{r},\bi{r}';W)=-\delta(\bi{r}-\bi{r}')
\label{eqn:gf3}
\end{equation}
for $\bi{r},\bi{r}' \in I$.
A similar line of argument to that given by Inglesfield \cite{inglesfield81}
for the embedded Schr\"odinger equation shows that this Green function is
identical for $\bi{r},\bi{r}' \in I$ to the Green functions 
$G_{\rm I+II}$ for the entire system I$+$II.
For simplicity assuming I$+$II constitute a finite system so that the 
spectrum is discrete, the Green function $G_{\rm I+II}$ is given by
\begin{equation}
G_{\rm I+II}(\bi{r},\bi{r}';W)=
\sum_n 
\frac{\Psi_n(\bi{r})\Psi_n^\dag(\bi{r}')
}{W-W_n^{\rm I+II}}
\label{eqn:gf4}
\end{equation}
where $W_n^{\rm I+II}$ is the eigenvalue corresponding to eigenstate
$\Psi_n(\bi{r})$ of the entire system, normalised to unity over I$+$II. 
For a given $W$, 
the Green function solving (\ref{eqn:gf3}) can be expanded in terms of 
the eigenstates $\varphi_n(\bi{r};W)$ of the corresponding homogeneous equation
\begin{equation}
\fl
\widehat{H}\varphi_n(\bi{r};W)+\int_I \rmd\bi{r}' \Delta(\bi{r},\bi{r}';W) 
\varphi_n(\bi{r}';W) -W_n(W)\varphi_n(\bi{r};W)=0 \qquad \bi{r}\in I
\label{eqn:gf5}
\end{equation}
normalised to unity over I, as
\begin{equation}
G(\bi{r},\bi{r}';W)=
\sum_n
\frac{\varphi_n(\bi{r};W)\varphi_n^\dag(\bi{r}';W)
}{W-W_n(W)}.
\label{eqn:gf6}
\end{equation}
Clearly $G$ has poles at $W=W_n(W)$. At these energies (\ref{eqn:gf5}) becomes
the exact embedded Dirac equation (\ref{eqn:gf1}) so 
as we have seen the poles will occur
at eigenstates of the entire system and the spectrum of $G$ and 
$G_{\rm I+II}$ coincide. It remains to show the poles of $G$
have the appropriate weight.  The residue of $G$ at $W_n$ is
\begin{equation}
\frac{\varphi_n(\bi{r};W_n)\varphi_n^\dag(\bi{r}';W_n)}{
1-\left.\left({\textstyle\partial W_n(W)}
{\textstyle\partial W}\right)\right|_{W_n}}
=
\frac{\varphi_n(\bi{r};W_n)\varphi_n^\dag(\bi{r}';W_n)}{
1-c^2\hbar^2\int_S \rmd\bi{r}_S\cdot 
\varphi^\dag_{\rm l}\bsigma 
\int_S \rmd\bi{r}_S'\cdot
\dot{\Gamma}(W_n) \bsigma\varphi_{\rm l}}
\label{eqn:gf7}
\end{equation}
The second term in the denominator is precisely the additional factor 
necessary to correctly normalise the states (see (\ref{eqn:norm}), 
(\ref{eqn:cgc})) so that
\begin{equation}
\frac{\varphi_n(\bi{r};W_n)\varphi_n^\dag(\bi{r}';W_n)}{
1-\left.\left({\textstyle\partial W_n(W)}/
{\textstyle\partial W}\right)\right|_{W_n}}
=\Psi_n^\dag(\bi{r})\Psi_n(\bi{r}').
\end{equation}
The residues of the Green function of the embedded system and those of the 
entire system are identical. Hence the two Green functions are
identical for $\bi{r},\bi{r}'\in $ I.

For practical calculations the Green function can be expanded
using a double-basis of separate large and small component spinors:
\begin{equation}
G(\bi{r},\bi{r}';W)=
\left[\begin{array}{cc}
\bpsi_{\rm l}(\bi{r}) & 0 \\ 0 & \bpsi_{\rm s}(\bi{r})
\end{array}\right]{G}(W)
\left[\begin{array}{cc}
\bpsi_{\rm l}(\bi{r}') & 0 \\ 0 & \bpsi_{\rm s}(\bi{r}')
\end{array}\right]^\dag.
\end{equation}
The matrix elements of the matrix of coefficients $\bi{G}$ may be found by
substituting into (\ref{eqn:gf3}), multiplying from the right by the 
vector of basis functions, multiplying from the left by the
Hermitian transpose of the vector of basis functions, and integrating over
region $I$. This leads to
\begin{equation}
{G}(W)
=\left[
\begin{array}{cc}
W{O}_{\rm ll}-{H}_{\rm ll} & -{H}_{\rm ls} \\
-{H}_{\rm sl} & W{O}_{\rm ss}-{H}_{\rm ss}
\end{array}
\right]^{-1}
\end{equation}
where the overlap and Hamiltonian matrices have their previous definitions
(\ref{eqn:h2}-\ref{eqn:h7}) with $\dot{\Gamma}=0$.

\begin{figure}
\bigskip

\epsfxsize= 80mm
\centerline{\epsffile{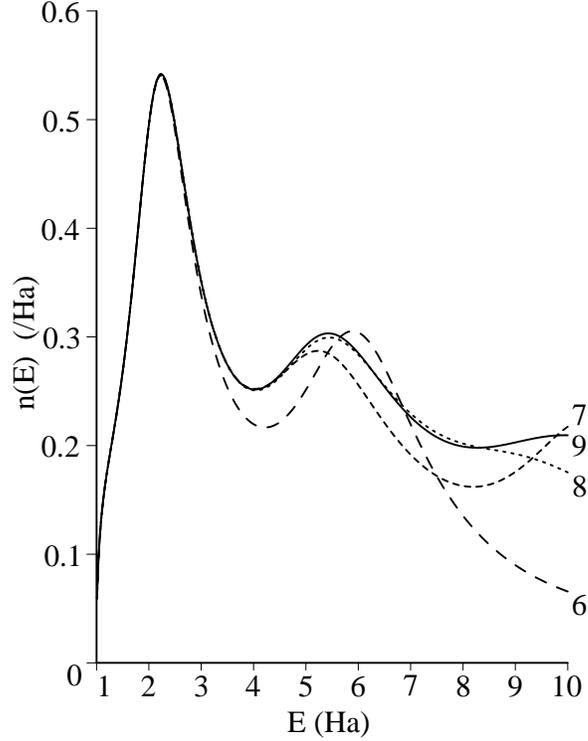}}
\bigskip

\caption{\label{fig:4}The $s_{1/2}$-wave local density of states as a function of 
energy $E=W-mc^2$, integrated through a sphere of
radius $R=3$ for the model system of a confined hydrogen 
atom with model parameters $R=3$, $V_0=1$. Different curves have been
calculated with basis sets corresponding to the indicated number 
$N_{\rm l}$ of basis functions.}
\end{figure}

As an illustration we calculate the local density of states for the confined
hydrogen model at energies above $V_0$ where the spectrum is continuous. 
Integrating over the embedded region this is given by
\begin{equation}
n(W)=-\frac{1}{\pi}{\rm Im}\ {\rm Tr}\ {G}(W+i0^+){O}.
\end{equation}
Figure \ref{fig:4} shows the $s_{1/2}$ wave local density of states 
for $R=3$, $V=1$, calculated with varying number of basis functions.
The basis functions (\ref{eqn:bas1}), (\ref{eqn:bas2}) are not particularly
appropriate for representing the continuum wave solutions, and so convergence
is only achieved using a relatively large set; however the results serve 
to illustrate the systematic improvement that accompanies an increasing number
of basis functions. The local density of states shows two resonances, the
precursors of bound states that exist when any of $R$, $Z$ or $V_0$ are
increased sufficiently.

\section{Application to an embedded monolayer}
\label{section:monolayer}
As a further example, one that provides a test of the relativistic
embedding scheme when applied
to a more challenging problem, we use it to calculate the local density
of states on a silver monolayer in a Au(001)/Ag/Au(001) sandwich structure.
Using the embedding scheme only the region occupied by the Ag monolayer 
is explicitly treated. This is region I, with the two Au halfspaces to 
either side
entering the calculation via embedding potentials expanded on planar surfaces.
Then, using Bloch's theorem the calculation is performed 
within a unit cell containing one atom. The full technical details will
be described elsewhere, but briefly the Green function at two-dimensional wave
vector $\bi{K}$ is expanded in a set 
of linearised augmented relativistic plane waves. We use large component
basis functions
\begin{equation}
\psi_{{\rm l},\bi{G}\sigma}(\bi{r})=
\left\{
\begin{array}{lll}
\varphi_\sigma
\exp(\rmi(\bi{K}+\bi{G})\cdot \bi{r}) && \bi{r} \in \mbox{interstitial} \\
\displaystyle \sum_\Lambda
\left[
A_{\kappa}^{\bi{G}\sigma}u_{\kappa}(r)+
B_{\kappa}^{\bi{G}\sigma}\dot{u}_{\kappa}(r)
\right]\Omega_\Lambda(\widehat{\bi{r}})
&& \bi{r} \in \mbox{muffin-tin}
\end{array}
\right.
\end{equation}
where $\varphi_\sigma$ is a Pauli spinor,
$\bi{G}=\bi{g}+G_z\hat{\bi{z}}$, with $\bi{g}$ a two-dimensional
reciprocal lattice vector 
and $G_z=n\times 2\pi/\tilde{D}$, $n=0,\pm 1, \pm 2, \dots$ and where 
$\tilde{D}$ exceeds the width of the embedded region
ensuring variational freedom in the basis. 
The function $u_{\kappa}$ is the 
large component of the wavefunction that satisfies the radial Dirac equation 
for the spherically symmetric component of the potential at some
pivot energy; $\dot{u}_{\kappa}$ is the energy derivative
of $u_\kappa$. The matching coefficients $A_{\kappa}^{\bi{G}}$, 
$B_{\kappa}^{\bi{G}}$ ensure continuity of the basis function
in amplitude and derivative at the muffin-tin radius. 
The small component basis functions are chosen to satisfy kinetic balance.

Overlap and Hamiltonian matrix elements follow directly from these basis
functions. The embedding potential is obtained from 
(\ref{eqn:cgc}) using the general expression for a wavefunction outside 
a surface at wave vector $\bi{K}$. This gives for the embedding potential
describing the left Au half space
\begin{eqnarray}
\fl
\Gamma(\bi{r}_S,\bi{r}'_S)=
\frac{\rmi}{W+mc^2}
\sum_{\bi{g}\sigma\bi{g}'\sigma'}
\left[
\left(S^-+S^+R^{+-}\right)\left(1+R^{+-}\right)^{-1}
\sigma_z\right]_{\bi{g}\sigma\bi{g}'\sigma'}\nonumber\\
\times
\exp(\rmi(\bi{K}+\bi{g})\cdot \bi{r}_S)
\exp(-\rmi(\bi{K}+\bi{g}')\cdot \bi{r}'_S)
\,\varphi_\sigma \otimes \varphi_{\sigma'}
\label{eqn:embpot}
\end{eqnarray}
with
\begin{equation}
S^\pm_{\bi{g}\sigma\bi{g}'\sigma'}=
\varphi_\sigma^\dag \left( \bsigma\cdot \bi{K}_{\bi{g}}^\pm
\right) \varphi_{\sigma'}\delta_{\bi{gg}'}
\end{equation}
The reflection matrix $R^{+-}$ is found using standard layer-scattering 
methods \cite{halilov93}. A similar approach may be used to obtain an
embedding potential for the right half space, which unlike the 
non-relativistic case differs from that for the left half space.

\begin{figure}
\bigskip

\epsfxsize=100mm
\centerline{\epsffile{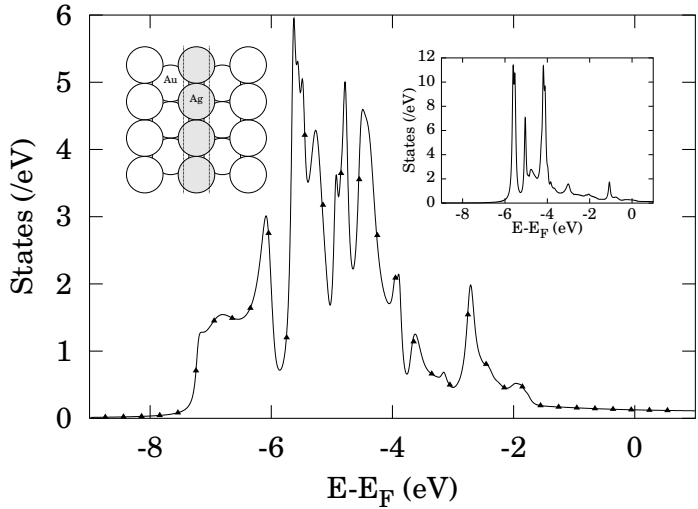}}
\bigskip

\caption{\label{fig:5}Local density of states on a Ag monolayer 
in a Au(001)/Ag/Au(001) sandwich structure calculated using a
relativistic scattering method (\full) and relativistic embedding
(\fulltriangle). The insets shows the calculation geometry (left) and
the local density of states obtained from the Schr\"odinger equation 
(right).}

\end{figure}

Figure \ref{fig:5} compares the local density of states calculated using
the relativistic embedding technique for an embedded Ag monolayer using
embedding potentials corresponding to Au(001)
with that found for an Au(001)/Ag/Au(001) sandwich geometry using relativistic
scattering theory \cite{halilov93}. The same Au and Ag potentials has been 
used in each 
case, and the
local density of states found within the same muffin-tin volume. Therefore
the results obtained with the two methods should be comparable, and we find 
that they are indistinguishable. This confirms that the embedding potential
(\ref{eqn:embpot}) imposes the correct variational constraint upon 
wave functions for the embedded Ag monolayer so that they replicate the 
behaviour of an extended Au(001)/Ag/Au(001) sandwich structure.
The inset in figure \ref{fig:5} shows the local density of states in the
non-relativistic limit ($c\longrightarrow\infty$), indicating the 
significant relativistic effects on the electronic structure which are 
correctly reproduced with this Dirac-embedding scheme.

\section{Summary and Discussion}
\label{section:summary}

We have outline above an embedding scheme for the Dirac equation. It enables
the Dirac equation to be solved within a limited region I when this region
forms part of a larger system, I$+$II. Region II is replaced by an additional
term added to the Hamiltonian for region I, and which acts on the surface
$S$ separating I and II. The embedding scheme is derived using a trial function
in which continuity in the small component across $S$ is imposed variationally.
Expanding the wave function in a basis set of separate large and small 
component spinors, the problem of variational collapse is avoided by 
using a basis satisfying kinetic balance. Calculating the spectrum of
a confined hydrogen atom, the method is shown to be stable and converge to
the exact eigenvalues.
We have also derived the Green function for the embedded Hamiltonian 
and illustrated its use in the continuum regime of the same confined hydrogen
system and an Au/Ag/Au sandwich structure. These are demonstration 
calculations -- future applications are likely to be 
to defects and surfaces of materials containing heavier (typically 5$d$)
elements, within the framework of density functional theory.

It is worthwhile to discuss further the use of a trial function that is 
discontinuous in the small component, since such a wave function gives rise to 
a discontinuous probability density and so would normally be dismissed
in quantum theory. In non-relativistic quantum mechanics
discontinuous trial functions are not permitted, since they possess
infinite energy. However the Dirac equation is first order in $\bi{p}$ and
as we have seen a perfectly regular expectation value of $H$ results. 
Exploiting this freedom, the embedding scheme outlined above leads to
solutions that are continuous in both large and small component {\it only}
when the embedding potential $\Gamma(\bi{r}_S,\bi{r}_S';w)$ is evaluated
at the same energy $w$ as the energy $W$ that appears in the Dirac equation 
itself, for then
the relationship between small and large components on $S$ inside 
(equation \ref{eqn:de2}) and outside (equation \ref{eqn:cgc}) coincide, 
the large components matching by construction. This may be achieved
for example via the iterative scheme used in connection 
with equation (\ref{eqn:exp3}) and the final column of table \ref{table:1}, 
or explicitly when determining the Green 
function as in Section \ref{section:green}. These are the methods in which
the non-relativistic embedding scheme has been most widely used.

When $w$ and $W$ do not coincide, the solutions obtained via this embedding
scheme will retain small components that are discontinuous across $S$.
This may be unacceptable for certain applications, but the solutions continue
to be valid approximations at least in as much as they provide estimates of 
the energies of the solutions of the Dirac equation, and so could
suffice e.g. for interpreting spectroscopic measurements. This embedding scheme 
places no greater 
emphasis on a discontinuity in the small amplitude at $S$ than on an incorrect 
(but continuous) amplitude elsewhere within the embedded region. It aims 
merely to optimise the
energy of the state, and will retain a discontinuity in the small component
if in doing so it can better (in terms of energy) approximate the solution 
inside the embedded region. In the non-relativistic embedding scheme
the discontinuous derivative of the trial function implies a probability
current (and electric current) that is discontinuous across the embedding 
surface. This is similarly unphysical, yet numerous applications such as 
those cited above have demonstrated the utility and accuracy of the method. 
Indeed, there have been many applications
in which this scheme has been used to determine currents and or transport 
properties, such as in relation to surface optical response 
\cite{optical} and electron transport in electron waveguides 
or through domain walls \cite{transport}. The reason for the
success of these calculations is that they employed schemes in which 
the embedding potential was evaluated at the correct energy, ensuring that
the derivative of the wavefunction was continuous across $S$. 
In practise there have been 
few calculations using the non-relativistic embedding scheme in which the
energies did not coincide.

There are a number of aspects of the method which are worthy of further 
consideration. We started with a trial function in which by construction
the large
component was continuous and the small component discontinuous 
across the surface $S$ dividing I and II. We could have
reversed these conditions, leading to a similar embedded Dirac equation
but with a modified embedding term. The particular
choice was motivated by the wish to have
a theory which behaves reasonably in the limit $c\longrightarrow\infty$ when 
the small component becomes negligible -- a discontinuous amplitude is 
not permissible in trial solutions to the Schr\"odinger equation. However, 
the behaviour of the alternative formulation should be investigated. Perhaps 
in connection with this there is the question of the spectrum of negative
energy solutions, to which we have paid scant attention.

Exploring the $c\longrightarrow\infty$ limit it might be possible to identify
how to embed a relativistic region I within a region II treated 
non-relativistically -- a 5$d$ overlayer on a simple metal substrate
might be a physical system where such a treatment is appropriate. 
There could be benefits in terms of computational resources expended
if the embedding potential could be determined
within the framework of a non-relativistic calculation, and there might 
also be useful insights in terms of simple models. Finally, 
in terms of implementation for realistic systems, some of the novel
schemes for deriving embedding potentials \cite{crampin92,ishida01} could 
certainly be adapted to the relativistic case. It would also be worthwhile 
to consider whether it is possible to use a restricted electron-like basis,
in which the large and small component spinors are combined. This is common
practice in most relativistic electronic structure calculations for 
solids when using basis set techniques (e.g. \cite{theileis00}), and
would result in significant computational efficiencies.

\end{document}